\documentstyle[preprint,prc,eqsecnum,epsfig,aps,floats,tighten]{revtex}

\def\Li{{\rm Li}}
\def\Re{{\rm Re}}
\def\half{{\textstyle{1\over 2}}}

\begin{document}
\draft
\preprint{JLAB-THY-02-59}

\title{Radiative corrections and parity-violating electron-nucleon
scattering}

\author{S.~Barkanova$^1$, A.~Aleksejevs$^1$, and P.G.~Blunden$^{1,2}$}
\address{$^1$Department of Physics and Astronomy, University of Manitoba,
Winnipeg, MB, Canada R3T~2N2}
\address{$^2$Jefferson Laboratory, 12000 Jefferson Avenue, Newport News,
VA 23606}

\maketitle

\begin{abstract}
Radiative corrections to the parity-violating asymmetry measured in
elastic electron-proton scattering are analyzed in the framework of the
Standard Model. We include the complete set of one-loop contributions to
one quark current amplitudes. The contribution of soft photon emission to
the asymmetry is also calculated, giving final results free of infrared
divergences. The one quark radiative corrections, when combined with
previous work on many quark effects and recent SAMPLE experimental data,
are used to place some new constraints on electroweak form factors of the
nucleon.
\end{abstract}


\narrowtext

\section{Introduction}

One of the areas of active theoretical and experimental interest is in
understanding the strange quark contribution to nucleon electroweak form
factors. The $s$ quarks are a key to the overall sea properties, and their
distribution is of particular interest for developing our understanding of
low energy nucleon structure. At present, theoretical estimates of
strangeness electric and magnetic form factors span a wide spectrum in
both magnitude and sign \cite{BM01}.

The electromagnetic current is a pure SU(3)$_f$ octet. By exploiting
flavor symmetries one can separate two flavor structures: the isovector,
$(u-d)$, and the hypercharge, $(u+d-2s)$. There is no singlet component
$(u+d+s)$, so the electromagnetic current can not provide sufficient
information to separate $u$, $d$, and $s$ contributions.

The $Z^0$ offers a new flavor coupling to the nucleon proportional to weak
isospin, which samples $(u-d-s)$ in the light quark sector.  Knowing
matrix elements of $(u-d)$, the $Z^0$ can be used as a probe to find
strange quark matrix elements in the nucleon \cite{KM88}. One can measure
such direct characteristics of strangeness in the nucleon as the strange
magnetic moment $\mu_s=\half s^\dagger\,\bbox{r}\times \bbox{\alpha}\,s$,
the strangeness charge radius $r_s^2=s^\dagger\,r^2\,s$, and the
strangeness analog of the axial charge
$\bar{s}\,\bbox{\gamma}\gamma_5\,s,$, which can serve as an independent
confirmation of the quark spin fraction measurements.

Electroweak properties of the nucleon can be studied by parity-violating
electron scattering at low to medium energies \cite{Bec89}. There the
asymmetry factor coming from the difference between cross sections of
left- and right-handed electrons can be measured.  However, extracting
information of interest from the experimental data requires evaluating
radiative corrections to electroweak scattering at the few percent level. 

Electroweak radiative corrections to intermediate energy, parity
non-conserving semi-leptonic neutral current interactions have been
addressed previously \cite{Mus89,MH90,MH91,Mus94,Zhu00}. In these works,
radiative corrections are constructed from the underlying fundamental weak
interaction between electron and quarks. Broadly, such corrections are
denoted as being either one-quark or many-quark effects. The one-quark
corrections involve the interaction of the electron with a single quark.
The many-quark contributions involve two or more quarks, and include
effects due to an intrinsic weak interaction in the nucleon (e.g. the
anapole moment). In this work, we restrict our considerations to the
one-quark contributions.

The use of computer packages $FeynArts$, $FormCalc$, and $LoopTools$
\cite{HP99} is of great assistance in facilitating our calculations,
allowing us to include the full set of one-loop contributions (there are
several hundred Feynman diagrams, which are laborious to calculate by
hand), and to retain the momentum-dependence of the amplitudes, for
example. Our treatment of hadronic model dependencies (e.g. kinematics)  
is slightly different than previous work. In addition, we have treated
infrared (IR) divergences in the one-loop amplitudes by including
bremsstrahlung contributions for soft photon emission. While such a
prescription is not completely satisfactory, and would be better handled
by also accounting for hard photons and the particular detector setup of a
given experiment, it nevertheless sets the scale for uncertainties of this
origin.

The article is described as follows. In Sec.~II we give the formalism for
parity-violating electron scattering, including definitions of the
relevant currents, couplings, form factors, and kinematics. Section III
outlines the one-loop calculations, including the prescription for
including soft photon emission. Results for the radiative corrections are
given in Sec.~IV. Finally, our results are combined with other
calculations of many-quark effects to discuss the implications for the
SAMPLE experiment.

\section{Formalism}

\subsection{Parity violating electron scattering}

At tree level, the electron-nucleon scattering amplitude consists of
two terms, ${\cal M}^\gamma$ (Fig.~1a) and ${\cal M}^Z$ (Fig.~1b).
\begin{figure}[ht]
\begin{center}
\epsfig{figure=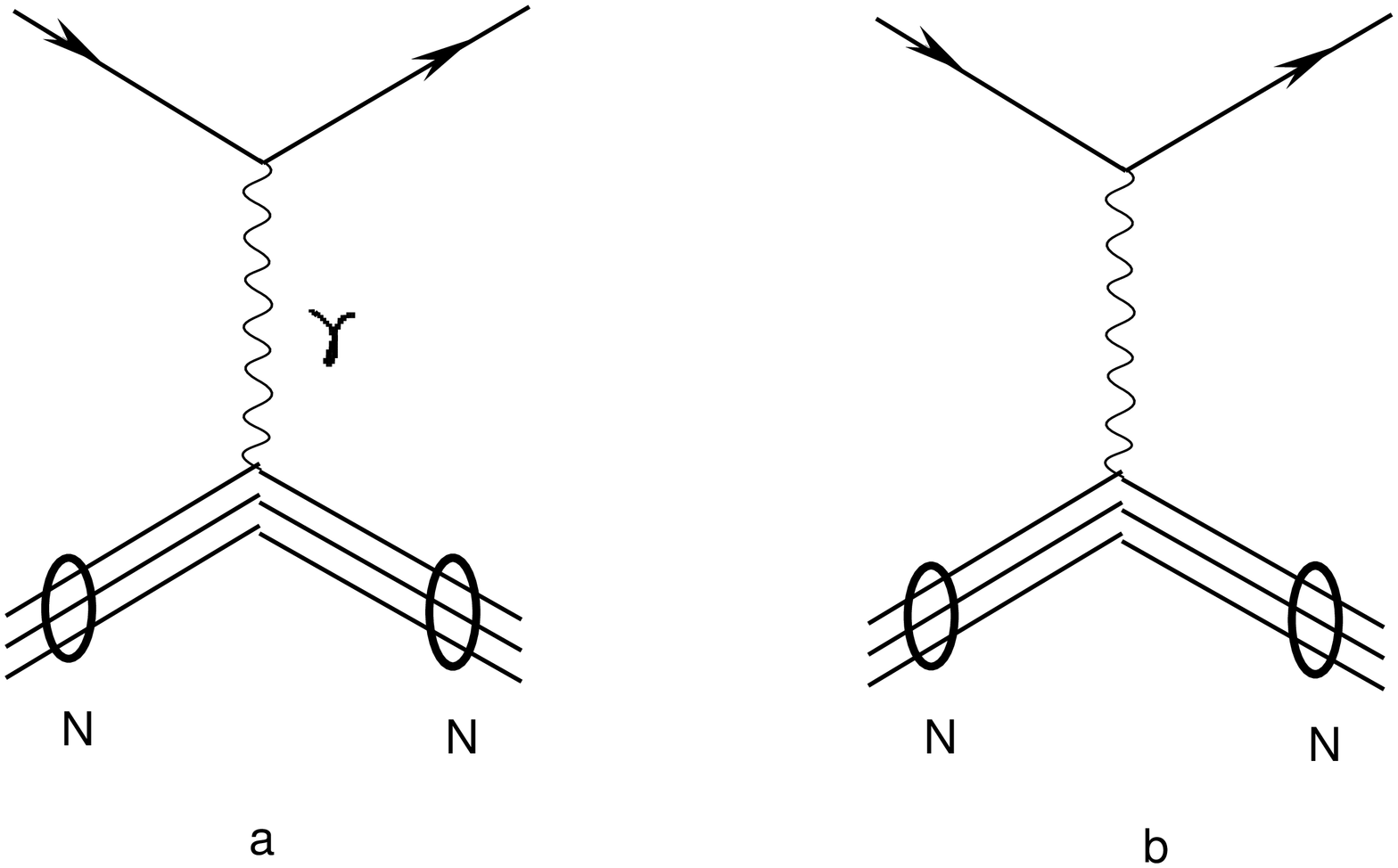,width=0.5\textwidth,
angle=270,clip=0}
\caption{Tree level electron-nucleon scattering.}
\end{center}
\end{figure}
The first term is parity-conserving $\gamma$-exchange, and the second has
a parity-violating contribution from $Z^0$-exchange. The differential 
cross section is proportional to
\begin{equation}  \label{e2.1}
d\sigma \sim \left| {\cal M}^\gamma\ +{\cal M}^Z\right| ^2=
\left|{\cal M}^\gamma\right| ^2
+2\Re\left\{\left({\cal M}^\gamma\right)^* {\cal M}^Z\right\}+\left|{\cal
M}^Z\right|^2.
\end{equation}
The purely weak term $\left|{\cal M}^Z\right|^2$ is very small compared
to the other terms and can safely be neglected. The
electromagnetic-weak interference term
$2\Re\left\{\left({\cal M}^\gamma\right)^* {\cal
M}^Z\right\}$ contains the physics of interest. This term can be
extracted from the parity-violating asymmetry, formed as the ratio of
helicity-dependent to helicity-independent cross sections: 
\begin{equation}
{\cal A}={d\sigma_R-d\sigma_L \over d\sigma_R + d\sigma_L}.
\end{equation}
Here, $\sigma_R$ and $\sigma_L$ are cross sections for right- and
left-handed electrons, respectively. This ratio, measured in a
number of experiments, is of the order 10$^{-6}$ at low energies.

For elastic electron-nucleon scattering, the asymmetry ratio is given
by \cite{Mus94,BH01}
\begin{eqnarray}  \label{e2.3}
{\cal A}&=&\left[ {-G_F Q^2\over 4\sqrt{2}\pi \alpha }\right] {\varepsilon
G_E^\gamma G_E^Z+\tau G_M^\gamma G_M^Z-(1-4\sin ^2\theta_W)\varepsilon
^{\prime }G_M^\gamma G_A^Z \over \varepsilon (G_E^\gamma )^2+
\tau (G_M^\gamma )^2}\nonumber\\
&\equiv& -{G_FQ^2 \over 4\sqrt{2}\pi \alpha }\times {N\over D},
\end{eqnarray}
where $Q^2>0$ is the four-momentum transfer. $G_E^\gamma $ and
$G_M^\gamma $ are the electric and magnetic vector form factors of the
nucleon associated with $\gamma$-exchange, $G_E^Z$ and $G_M^Z$ are the
similar parameters for $ Z^0$-exchange, and $G_A^Z$ is the axial vector
form factor. Kinematic parameters $\tau$, $\varepsilon$ and
$\varepsilon ^{\prime }$ are defined as 
\begin{eqnarray}
\tau &=&\frac{Q^2}{4M_N^2},  \nonumber \\
\varepsilon &=&\frac 1{1+2(1+\tau )\tan ^2\frac \theta 2},  \label{e2.4} \\
\varepsilon ^{\prime } &=&\sqrt{\tau (1+\tau )(1-\varepsilon ^2)}. \nonumber
\end{eqnarray}

The electromagnetic vector form factors of the proton $G_E^{\gamma ,p}$
and $G_M^{\gamma ,p}$ are well measured, and some constraints can be put
on $G_E^{\gamma ,n}$ and $G_M^{\gamma ,n}$ for the neutron \cite{Bra02}.  
The expressions for the weak form factors of the nucleon, $G_E^Z$ and
$G_M^Z$, and the axial vector coupling $G_A^Z$, are given below. The
factor $(1-4\sin ^2\theta_W)\simeq 0.1$ suppresses the term containing
$G_A^Z$, and makes the contribution of higher order processes more
apparent.

The neutral weak vector form factors for the proton and neutron can be
expressed in terms of the electromagnetic form factors of the proton
and neutron, plus a contribution from strange quarks \cite{Bei96}. We
choose to write this in the form
\begin{eqnarray}
G_{E,M}^Z&=&(1-2\sin^2\theta_W)\left[ 1+R_V^{T=1}\right]
G_{E,M}^{T=1}\tau_3\nonumber\\
&&-2\sin^2\theta_W \left[ 1+R_V^{T=0}\right]
G_{E,M}^{T=0}-\left[1+R_V^{(0)}\right] G_{E,M}^s. \label{e2.5} 
\end{eqnarray}
Here $R_V^{T=0}$, $R_V^{T=1}$, and $R_V^{(0)}$ are the isoscalar,
isovector, and isosinglet weak radiative corrections describing the
contribution from weak vector couplings beyond tree level, respectively.
Strong isospin $\tau_3$ is equal to $+1(-1)$ for the proton (neutron). The
isoscalar and isovector electromagnetic form factors are taken to be the
linear combinations
\begin{eqnarray}
G_{E,M}^{T=0} &=& G_{E,M}^p + G_{E,M}^n,\nonumber\\
G_{E,M}^{T=1} &=& G_{E,M}^p - G_{E,M}^n,
\end{eqnarray}
so that at $Q^2=0$, $G_E^{T=0}(0)=G_E^{T=1}(0)=1$, and
$G_M^{T=0}(0)=0.8797$, $G_M^{T=1}(0)=4.709$. The strange quark form
factors are undetermined, but take on the values $G_E^s(0)=0$ and
$G_M^s(0)=\mu_s$, where $\mu_s$ is the contribution of strange quarks to
the magnetic moment of the nucleon.

The axial vector coupling is conveniently written in terms of octet matrix
elements \cite{Mus94} as
\begin{equation}  \label{e2.6}
G_A^Z=-\left[ 1+R_A^{T=1}\right] G_A^{T=1}\tau_3+\sqrt{3}
R_A^{T=0}G_A^{(8)}+\left[ 1+R_A^{(0)}\right] G_A^s.
\end{equation}
The isovector axial form factor $G_A^{T=1}$ is determined from neutron
beta decay as $G_A^{T=1}(0)= g_A$, with $g_A=1.2670\pm 0.0035$. The second
term involving the $SU(3)$ isoscalar octet form factor $G_A^{(8)}$ is not
present at tree level. From a least squares fit to hyperon beta decay, we
have determined $G_A^{(8)}=0.169\pm0.009$, which is consistent with
results of polarized deep inelastic lepton scattering \cite{FJ01}. In our
calculations (see section IV) we find $|R_A^{T=0}|<|R_A^{T=1}|$, so we
expect overall that the second term in Eq.~(\ref{e2.6}) is suppressed
relative to the first term. The axial strange form factor $G_A^s$ is
extracted from polarized deep inelastic lepton scattering as
$G_A^s(Q^2=0)=-0.086\pm 0.024$ \cite{FJ01}. Due to the unknown $Q^2$
dependence we take $G_A^s(Q^{2}=0.1)=-0.086\pm 0.086$ in our analysis.

Additional information on the axial vector form factor can be provided
by parity-violating quasielastic scattering from deuterium. In the
simplest impulse approximation, the asymmetry ratio for the deuteron can be
written as the incoherent sum of neutron and proton contributions
\begin{equation}  \label{e2.7}
{\cal A}_d=\left[ {-G_F Q^2\over 4\sqrt{2}\pi \alpha }\right]
{N_n+N_p \over D_n+D_p},
\end{equation}
where $N_p (N_n)$ is the numerator and $D_p (D_n)$ is the denominator
in Eq.~(\ref{e2.3}) for the proton (neutron).

\subsection{Radiative effects}

As was indicated in the previous section, extracting form factors
$G_{E,M}^Z$ and $G_A^Z$ is based not only on experimental data, but also
on theoretical calculations of radiative corrections for parity-violating
scattering. We calculate the radiative corrections to tree level for the
electroweak interaction in the scattering processes $e+p\rightarrow e+p$
and $e+n\rightarrow e+n$ by modeling the nucleon as a collection of
quasi-free quarks carrying some fraction $x$ of the nucleon four-momentum.

If we assume that during the scattering process the electron interacts
with only one quark, we can split the problem of electron-nucleon
scattering into calculations of Feynman graphs for processes like
$e(p_1)+q(p_2)\rightarrow e(p_3)+q(p_4)$, with $q$ representing
$(u,d,s)$ quarks. Starting from the fundamental coupling of an
elementary fermion to the photon or to the $Z^0$, one can arrive at the
general form of electromagnetic and weak invariant amplitudes
\cite{Mus94}:
\begin{mathletters}
\begin{eqnarray}
{\cal M}^\gamma  &=&-{4\pi \alpha \over q^2}Q_f l^\mu J_\mu^\gamma, \\
{\cal M}^Z &=&-{4\pi \alpha\over M_Z^2-q^2}
{1\over (4 \sin\theta_W \cos\theta_W)^2}
\left(g_V^f l^\mu +g_A^f l^{\mu
5}\right)\left(J_\mu^Z+J_{\mu 5}^Z\right),
\end{eqnarray}
\end{mathletters}
where $l^\mu$ ($l^{\mu 5}$) and $J^\mu$ ($J^{\mu 5}$) are leptonic
and hadronic vector (axial vector) currents, respectively, and $Q_f$
is the electromagnetic charge number of the fermion.

The weak coupling has been expressed in terms of the set $(\alpha, M_W,
M_Z)$ rather than $G_F$, noting that $\sin^2\theta_W=1-M_W^2/M_Z^2$ in
our renormalization scheme. The vector and axial-vector ``charges'' of
the fermion, $g_V^f$ and $g_A^f$, are defined as
\begin{mathletters}
\begin{eqnarray}
g_V^f &=&2T_3^f-4Q_f\sin ^2\theta _W, \\
g_A^f &=&-2T_3^f,
\end{eqnarray}
\end{mathletters}
with $T_3^f=+\half(-\half)$ for the upper (lower) member of the
fermion doublet.

Explicitly, we have for the electron and the $(u,d,s)$ quarks:
\begin{eqnarray}
g_V^e &=&-1+4\sin^2\theta_W, \qquad g_A^e =+1, \nonumber\\
g_V^u &=&+1-{8\over 3}\sin^2\theta_W, \qquad g_A^u =-1,  \label{e1.cde} \\
g_V^{d,s}&=&-1+{4\over 3}\sin ^2\theta _W, \qquad g_A^{d,s} = +1.\nonumber
\end{eqnarray}
For the case of electron scattering, leptonic vector and axial-vector
currents are Dirac currents with electron spinor $u_e$:
\begin{mathletters}
\begin{eqnarray}
l^\mu  &=&\overline{u}_e\gamma ^\mu u_e, \\
l^{\mu 5} &=&\overline{u}_e\gamma ^\mu \gamma ^5u_e. 
\end{eqnarray}
\end{mathletters}
Hadronic currents $J_\mu$ are hadronic matrix elements of the
electro-magnetic, vector, and axial-vector quark current operators: 
\begin{equation}
J_\mu=\langle N\left| \widehat{J}_\mu\right| N\rangle,
\end{equation}
($J_\mu=J_\mu ^\gamma ,\ J_\mu ^Z$ or $J_{\mu 5}^Z$ ; $\left| N\right\rangle
=\left| p\right\rangle $ or $\left| n\right\rangle$),
\begin{eqnarray}
\widehat{J}_\mu ^\gamma  &=&\sum_qQ_q\overline{u}_q\gamma _\mu u_q,  \nonumber
\\
\widehat{J}_\mu ^Z &=&\sum_q{ g}_V^q\overline{u}_q\gamma _\mu u_q,
\label{e1.h} \\
\widehat{J}_{\mu 5}^Z &=&\sum_q{ g}_A^q\overline{u}_q\gamma _\mu \gamma
_5u_q.  \nonumber
\end{eqnarray}

Nominally, summation must be done over all quark flavors, but it is
sufficient at the momentum scale of interest to include only light quarks,
$u$, $d$, and $s$.  In order to be parity-violating, $Z^0$ exchange must
involve either $V(e)\times A(q)$ or $A(e)\times V(q).$ Thus, it is
convenient to express the amplitude for the parity-violating (PV) part of
the electron-quark $(eq)$ scattering in the general form
\begin{eqnarray}
{\cal M}_{\rm PV}^{eq} &=&A^{eq}\,\bar u_e(p_3)
\gamma^\mu u_e(p_1) \cdot \bar u_q(p_4)
\gamma_\mu \gamma_5 u_q(p_2) \nonumber \\
&&\quad +B^{eq}\, \bar u_e(p_3) \gamma^\mu \gamma_5 u_e(p_1)\cdot
\bar u_q(p_4) \gamma_\mu u_q(p_2),\label{e1.1} \\
&\equiv& A^{eq}\,J_{VA}^{eq}\ +\ B^{eq}\,J_{AV}^{eq}. 
\nonumber
\end{eqnarray}
Here, $A^{eq}$ and $B^{eq}$ are functions of the Standard Model
parameters, $n-$point tensor coefficients, and kinematics. From this point
we define one-loop radiative correction to tree level electron-quark
scattering as follows: 
\begin{equation}
R_{A}^{q}\equiv {A_{\rm rad}^{eq} \over A_{\rm tree}^{eq}},\qquad
R_{V}^{q}\equiv {B_{\rm rad}^{eq} \over B_{\rm tree}^{eq}},
\label{e1.2}
\end{equation}
where $A_{\rm tree}^{eq}$ and $B_{\rm tree}^{eq}$ come from the tree
level, and $A_{\rm rad}^{eq}$ and $B_{\rm rad}^{eq}$ come from the
one-loop radiative corrections. The relationship between the hadronic
radiative corrections of Eqs.~(\ref{e2.5}-\ref{e2.6}) and the
electron-quark radiative corrections of Eq.~(\ref{e1.2}) follows from the
linear combinations:
\begin{eqnarray}
R_V^{T=0} &=& (g_V^u R_V^u + g_V^d R_V^d)/(g_V^u + g_V^d), \nonumber\\
R_V^{T=1} &=& (g_V^u R_V^u - g_V^d R_V^d)/(g_V^u - g_V^d), \nonumber\\
R_V^{(0)} &=& (g_V^u R_V^u + g_V^d R_V^d + g_V^s R_V^s)/
(g_V^u + g_V^d + g_V^s), \\
R_A^{T=0} &=& - R_A^u + R_A^d, \nonumber\\
R_A^{T=1} &=& (R_A^u + R_A^d)/2, \nonumber\\
R_A^{(0)} &=& -R_A^u + R_A^d + R_A^s. \nonumber
\end{eqnarray}
For the axial corrections we have used the explicit values of $g_A^q$,
which are $+1$ or $-1$, and the $R_A^{T=0}$ correction is calibrated to
1, since the tree level term $g_A^u + g_A^d=0$.

Each particle carries 4-momentum $p_i^2=m_i^2$ with the following
structure in the center-of-mass (CM) frame:
\begin{eqnarray}
p_1 &=&\left( E_1,0,0,p\right) ,  \nonumber  \\
p_2 &=&\left( E_2,0,0,-p\right) ,  \nonumber \\
p_3 &=&\left( E_3,p \sin\theta ,0,p \cos\theta \right), \\
p_4 &=&\left( E_4,-p\sin\theta ,0,-p\cos\theta \right).
\nonumber
\end{eqnarray}
Here $E_1=E_3$ and $E_2=E_4$ (for elastic scattering) and $p$ denote
the energy and momentum of the scattered particles. The Mandelstam
variables are defined as follows:
\begin{eqnarray}
s &=&\left( p_1+p_2\right) ^2=\left( E_1+E_2\right) ^2=E_{\rm CM}^2,
\nonumber \\
t &=&\left( p_1-p_3\right) ^2=-2p^2\left( 1-\cos\theta \right),
\label{e1.4} \\
u &=&\left( p_1-p_4\right) ^2=\left( E_1-E_2\right) -2p^2\left( 1+\cos
\theta \right),   \nonumber
\end{eqnarray}
with $E_{1,2}^2=p^2+m_{1,2}^2$. For elastic scattering we have
$s+t+u=2m_1+2m_2$. Input kinematic parameters, in our case, are the energy
of the electron in the laboratory reference frame, $E_{\rm lab}$, and
the scattering angle $\theta$. Hence $E_{\rm CM}$ and $p$ can be expressed
in terms of $E_{\rm lab}$ and $\theta $:
\begin{equation}
p^2=\frac{\left( E_{\rm CM}^2+m_2^2-m_1^2\right) ^2}{2E_{\rm CM}^2}-m_2^2.
\end{equation}
For quarks bound in the nucleon, we can assume that the energy of the
quark can be defined as a fraction of the mass of the nucleon
$E_{2,{\rm lab}}^2=x^2 m_N^2$, with $x\approx 1/3$. Taking into account
$E_{2,{\rm lab}}^2=p_{2,{\rm lab}}^2+m_{2}^2$, we can determine
center-of-mass energy as follows:
\begin{eqnarray}
E_{\rm CM}^2 &=&\left( p_{1,{\rm lab}}+p_{2,{\rm lab}}\right) ^2 \nonumber\\
&=&m_{1}^2+m_{2}^2+2E_{1,{\rm lab}}\,x\,m_N \left(1-{(E_{1,{\rm lab}}^2-
m_{1}^2)^{1/2} ((x\,m_N)^2-m_{2}^2)^{1/2}
\cos\theta_i \over E_{1,{\rm lab}}\,x\,m_N}\right).
\end{eqnarray}
In the latter equation, $\theta_i$ denotes the unknown angle between the
spatial momentum of the electron and quark just before scattering took
place. Final expressions for radiative corrections have been integrated
over this angle to account for all possible values of $\theta_i.$

\section{Details of the calculation}
\subsection{One-loop corrections}

Doing one-loop calculations in the Standard Model by hand is a laborious
task due to the sheer number of particles and the complexity of the
underlying theory. To facilitate this task, and to reduce the possibility
of errors, for some time software packages have been developed to automate
different tasks in these calculations \cite{Hah00}. In this paper, we have
chosen to use the packages {\it FeynArts}, {\it FormCalc}, and {\it
LoopTools} \cite{HP99}. These packages are designed to work hand in hand
on various aspects of the calculation, as described below.
 
In total, 438 one-loop diagrams were calculated, including counterterms
and contributions from scalar bosons $\left( H,\chi ,\phi \right)$ and
ghost fields $\left(u_{Z},u_{-},u_{+}\right)$. For analysis and comparison
with earlier work (see Ref.~\cite{Mus89}, for example), it is useful to
split diagrams into three classes (Fig.~2).  Class 1 are self-energy
loops, Class 2 are triangles, and Class 3 are boxes. Class 1 have a
dominant contribution from $Z$ exchange and $\gamma Z$ mixing. Class 2
include triangles with $ff'Z$ and $ff'\gamma$ vertices. Class 3 are the $Z
Z$, $W W$ and $\gamma Z$ box and crossed box diagrams.  Contributions from
scalar bosons and ghost fields to the boxes were found to be negligibly
small. 

\begin{figure}[ht]
\begin{center}
\epsfig{figure=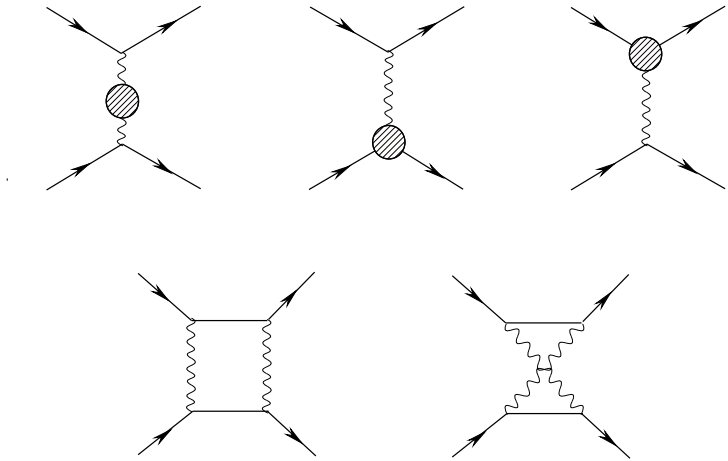,width=0.5\textwidth,
angle=270, clip=0}
\caption{Classes of one-loop contributions to electron-quark scattering
include self-energy, vertex, and box diagrams.}
\end{center}
\end{figure}

Because of the huge size of the analytical expressions for amplitudes, we
leave them out of this article. Certain steps involved in the evaluation
of one-loop PNC amplitude have to be explained more carefully. To evaluate
four dimensional one-loop tensor integrals, we have used the Constrained
Differential Renormalization (CDR) scheme \cite{Agu98} implemented in the
package {\it FormCalc}. In the CDR scheme, Feynman diagrams are considered
completely in four dimensions.  Thereafter, the reduction of singular
basic functions (products of propagators and their derivatives) has been
renormalized into the sum of ``regular'' ones by implementing a set of
rules in such a way that Ward identities are satisfied. It was proven in
Ref.~\cite{HP99} that CDR is equivalent at the one-loop level to
regularization by dimensional reduction \cite{Sie79}, after taking the
Fourier transform of the basic renormalized functions into momentum space.
This last approach corresponds to a modified dimensional regularization,
where one-loop integrals are considered in $D$ dimensions, but all the
tensors and spinors are kept 4-dimensional \cite{Sie79}. To preserve gauge
invariance in dimensional reduction, one should use $\widehat{g}_{\mu
\nu}\,\left( \widehat{g}_\mu ^\mu =D\right)$ with g$_{\mu \nu }\widehat{g}
^{\nu \rho }=\widehat{g}_\mu ^\rho $ for the tensor decomposition. 

In dimensional regularization, the general structure of one-loop tensor
integral can be written in the form \cite{HS99}
\begin{equation}
J_{i_1\ldots i_p}^N={\left( 2\pi \mu \right) ^{4-D} \over i\pi ^2}
  \int d^Dq\,{q_{i_1}q_{i_2}\ldots q_{i_p} \over \left( q^2-m_1^2\right)
    ( \left(q+p_1\right)^2-m_2^2) \cdots 
    ( \left( q+p_{N-1}\right)^2-m_N^2 )}.  \label{e1.8}
\end{equation}
Here, $(\mu) $ means the regularization scale parameter of dimensional
reduction, which is related to the CDR renormalization scale by
$\log\left({\bar M}^2\right) =\log\left(\mu^2\right)+2$. For the tensor
decomposition of Eq.~(\ref{e1.8}) into linear combinations of tensor
coefficients functions, we have used the method of Passarino-Veltman
\cite{PV79}. 

The Passarino-Veltman approach deals with two, three, and four point
tensor integrals with two, three, and four propagators, respectively.
The rank of the tensors is equal to the number of integrable momenta
$\left(q_{i_1}q_{i_2}\ldots q_{i_p}\right) $in the numerator of
Eq.~(\ref{e1.8}). Two, three and four point tensor coefficient functions,
coming from tensor decomposition, were calculated numerically by using the
package {\it LoopTools}, and algebraic calculations were completed with
the help of packages {\it FeynArts}, {\it FormCalc} and {\it FORM}. To
summarize this section, we provide an outline of how calculations were
implemented in these packages.
\\~\\
Generation of the diagrams (package {\it FeynArts})
\begin{itemize}
\item Topologies are defined
\item Fields are inserted
\item Amplitudes and their counterterms in integral form were defined,
along with CKM matrix
\end{itemize}
Evaluation of amplitudes (package {\it FormCalc} along with {\it FORM})
\begin{itemize}
\item Indices contracted and traces are taken
\item Amplitudes in general form (i.e. combination of many-point tensor
coefficients along with spinor chains) were presented
\item Radiative corrections were calculated (On Shell Renormalization)
\item Infrared divergences are treated by adding soft-photon emission
\end{itemize}
Numerical evaluation
\begin{itemize}
\item Standard Model parameters and kinematics are defined
\item Many-point tensor coefficients are numerically evaluated (package
{\it LoopTools})
\end{itemize}

\subsection{Renormalization constants}

Generally, tensor coefficient functions are ultraviolet divergent
(inversely proportional to the parameter $\varepsilon =4-D$).  In order to
cancel divergences and transform bare parameters into physical observables
one has to introduce a renormalization scheme.  The renormalized
parameters are related to the bare parameters (denoted by a subscript 0)
as follows: 
\begin{eqnarray}
 M_{Z,0}^2 &=& M_Z^2 + \delta M_Z^2,\nonumber\\
 M_{W,0}^2 &=& M_W^2 + \delta M_W^2,\nonumber\\
 M_{H,0}^2 &=& M_H^2 + \delta M_H^2,\nonumber\\
 m_{f_i,0} &=& m_{f_i} + \delta m_{f_i},\nonumber\\
 e_0 &=& (1+\delta e)\,e,\nonumber\\
 \left( \begin{array}{c}
   Z_0 \\
   A_0
 \end{array} \right) &=&
 \left( \begin{array}{cc}
   1+\half\delta Z^{ZZ} & \half\delta Z^{ZA}\\
   \half \delta Z^{AZ} & 1+\half\delta Z^{AA}
 \end{array} \right)
 \left( \begin{array}{c}
   Z \\
   A
 \end{array} \right), \\
 W^{\pm}_0 &=& \left( 1+\half\delta Z^{WW} \right)\,W^\pm,\nonumber\\
 H_0 &=& \left( 1+\half\delta Z^H \right)\,H,\nonumber\\
 f^L_{i,0} &=& \left( \delta_{ij}+\half\delta Z^{f,L}_{ij} \right)
   \,f^L_j, \nonumber\\
 f^R_{i,0} &=& \left( \delta_{ij}+\half\delta Z^{f,R}_{ij} \right)
   \,f^R_j. \nonumber
\end{eqnarray} 
Counterterms were chosen in the On Shell Renormalization (OSR) scheme
in the 't~Hooft-Feynman gauge, where the gauge parameter $\xi =1$.
The renormalization constants are \cite{Hah97}:
\\~\\
Wave function renormalization:
\begin{eqnarray}
\delta Z^{ZZ} &=& -\Re \left( {\partial \over \partial k^2}
  \Sigma_\perp^{ZZ}\left( k^2\right) \right)_{k^2=M_Z^2},\qquad
\delta Z^{ZA} = 2\Re \left( {\Sigma_\perp^{AZ}\left(0\right)
  \over M_Z^2}\right), \nonumber\\
&& \nonumber\\
\delta Z^{AA} &=& -\Re \left( { \partial \over \partial k^2}
  \Sigma_\perp^{AA}\left( k^2\right) \right)_{k^2=0},\qquad
\delta Z^{AZ} = -2 \Re \left( {\Sigma_\perp^{AZ}\left( M_Z^2\right)
  \over M_Z^2}\right), \nonumber\\
&&  \nonumber \\
\delta Z^{WW} &=& -\Re \left( {\partial \over \partial k^2}
  \Sigma_\perp^{WW}\left( k^2\right) \right)_{k^2=M_W^2},\qquad
\delta Z^H = -\Re \left( {\partial \over \partial k^2}
  \Sigma^H\left( k^2\right)\right)_{k^2=M_H^2}, \\
\delta Z^\chi &=& -\Re \left( {\partial \over \partial k^2}
  \Sigma^\chi \left( k^2\right) \right)_{k^2=M_Z^2},\qquad
\delta Z^\phi = -\Re \left( {\partial \over \partial k^2}
  \Sigma^\phi \left( k^2\right) \right)_{k^2=M_W^2}, \nonumber\\
&& \nonumber\\
\delta Z_{ii}^{f,L} &=& -\Re \left(\Sigma_{ii}^{f,L}
  \left(m_{f_i}^2\right) \right) \nonumber \\
  && -m_{f_i}^2 \Re \left( {\partial \over \partial p^2}\left[
    \Sigma_{ii}^{f,L}\left(p^2\right) + \Sigma_{ii}^{f,R}\left(p^2\right)
    + 2\Sigma_{ii}^{f,S}\left(p^2\right) \right] \right)_{p^2=m_{f_i}^2}, \nonumber\\
&& \nonumber\\
\delta Z_{ii}^{f,R} &=& -\Re \left(\Sigma_{ii}^{f,R}
  \left(m_{f_i}^2\right) \right) \nonumber \\
  && -m_{f_i}^2 \Re \left( {\partial \over \partial p^2}\left[
    \Sigma_{ii}^{f,L}\left(p^2\right) + \Sigma_{ii}^{f,R}\left(p^2\right)
    + 2\Sigma_{ii}^{f,S}\left(p^2\right) \right] \right)_{p^2=m_{f_i}^2}. \nonumber
\end{eqnarray}

Mass renormalization:
\begin{eqnarray}
\delta M_Z^2 &=& \Re \left( \Sigma_\perp^{ZZ}\left( M_Z^2\right)
  \right), \qquad
\delta M_W^2 = \Re \left( \Sigma_\perp^{WW}\left( M_W^2\right)
  \right), \\
\delta M_H^2 &=& \Re \left( \Sigma^H\left( M_H^2\right) \right),\quad
\delta m_{f_i} = \half m_{f_i} \Re \left( \Sigma_{ii}^{f,L}\left( m_{f_i}^2\right)
  + \Sigma_{ii}^{f,R}\left( m_{f_i}^2\right)
  + 2\Sigma_{ii}^{f,S}\left( m_{f_i}^2\right) \right). \nonumber
\end{eqnarray}
Here $L$ and $R$ correspond to left- and right-handed fermions,
$\Sigma$ means one-loop integral of the truncated self-energy graph,
and $\perp$ denotes the transverse part only.

Charge and mixing angle renormalization:
\begin{eqnarray}
\delta \sin^2\theta_W &=&\cos^2\theta_W\left(\, {\delta M_Z^2 \over M_Z^2}-
  {\delta M_W^2 \over M_W^2}\right), \qquad
\delta \cos^2\theta_W\,=-\delta \sin^2\theta_W,  \\
\delta e &=& -{1\over 2} \left( \delta Z^{AA}+
{\sin^2\theta_W \over \cos^2\theta_W}\,\delta Z^{ZA}\right), \nonumber
\end{eqnarray}
with $\sin^2\theta_W=1-M_W^2/M_Z^2$.

\subsection{Infrared divergences and soft-photon emission}

For diagrams with photon exchange between external fermion legs we have
encountered infrared (IR) divergences. These are regulated by introducing
a small photon ``rest mass'' $\lambda$ in loop integrals involving the
photon propagator. This results in $\ln \left( m_f^2/\lambda ^2\right)$
terms in the amplitude. Hence the radiative corrections are, strictly
speaking, infinite in the limit $\lambda\rightarrow 0$. This unphysical
dependence on $\lambda$ is cancelled by adding the inelastic
bremsstrahlung contributions arising from soft-photon emission by the
external fermions (i.e. from the initial or final electron or quark) to
the scattering cross section \cite{BN37}. 

To see how this works in the parity-violating asymmetry, consider the
inelastic bremsstrahlung amplitude ${\cal M}^Z_{\rm b}$, which includes
the four diagrams of Fig.~3 involving $Z$ exchange, and the corresponding
QED amplitude ${\cal M}^\gamma_{\rm b}$ (not shown)  involving photon
exchange. The total inelastic cross section is given by
\begin{equation}  \label{dsiginel}
d\sigma_{\rm b}^{eq} \sim \left| {\cal M}^\gamma_{\rm b}\ 
+{\cal M}^Z_{\rm b}\right|^2.
\end{equation}
In analogy with Eq.~(\ref{e2.1}), a parity-violating asymmetry will
result from the interference term
$2\Re\left\{\left({\cal M}^\gamma_{\rm b}\right)^* {\cal M}^Z_{\rm b}\right\}$.

\begin{figure}[ht]
\begin{center}
\epsfig{figure=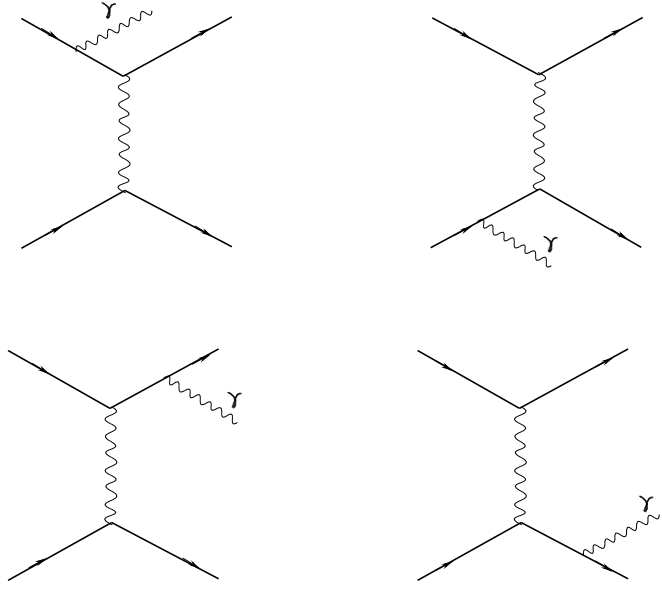,width=0.5\textwidth,
angle=270, clip=0}
\caption{Bremsstrahlung diagrams for treating IR divergences.}
\end{center}
\end{figure}

In the soft photon approximation, it is straightforward to show (see
Ref.~\cite{MT00}, for example) that the amplitude for soft photon
emission can be written in terms of the Born (tree level) amplitude
\begin{equation}
{\cal M}^\gamma_{\rm b} = {\cal M}^\gamma_{\rm Born} \left[
  -{p_1 \over\left( p_1\cdot k\right) }
  +{p_3 \over\left( p_3\cdot k\right) }
  +{Q_q p_2 \over\left( p_2\cdot k\right) }
  -{Q_q p_4 \over\left( p_4\cdot k\right) }
\right]\cdot \epsilon^*(k),\label{mgammasoft}
\end{equation}
with $k$ the momentum of the photon and $\epsilon(k)$ its polarization.
This is just the amplitude for elastic scattering (without bremsstrahlung)
times a factor for soft photon emission. An identical factor relates
${\cal M}^Z_{\rm b}$ to ${\cal M}^Z_{\rm Born}$. 

The contribution of the parity-violating inelastic differential cross
section to the radiative correction can also be expressed in terms of
the parity-violating elastic Born cross section, once the appropriate
integrals over phase space are done. The steps involved in carrying out
the phase space integrals are analogous to ordinary parity-conserving
electron scattering, as discussed for example in the paper by Maximon
and Tjon \cite{MT00}. We refer the reader to that paper for details,
and simply present the result that
\begin{equation}
d\sigma_{\rm b}^{eq} = d\sigma_{\rm Born}^{eq} \kappa_{\rm
soft}^{eq},
\end{equation}
where $\kappa_{\rm soft}^{eq}$ is the soft-photon factor defined as
\begin{equation}
\kappa_{\rm soft}^{eq}=-{\alpha \over 2\pi^2}\int_{\left| \bbox{k}\right|
  \leq \Delta \varepsilon}{d^3k \over 2\sqrt{\bbox{k}^2+\lambda ^2}}
\left(
  -{p_1 \over\left( p_1\cdot k\right) }
  +{p_3 \over\left( p_3\cdot k\right) }
  +{Q_q p_2 \over\left( p_2\cdot k\right) }
  -{Q_q p_4 \over\left( p_4\cdot k\right) }
\right)^2. \label{e1.15}
\end{equation}
Here $\Delta \varepsilon$ is the maximum momentum of the undetected
photon for which an elastic scattering event is recorded. It is related
to the final electron detector acceptance in the lab frame, $\Delta E$.

Adding together the elastic and inelastic cross sections, we have
\begin{equation}\label{e1.15b}
d\sigma_{\rm tot}^{eq} \sim
\left| {\cal M}^\gamma_{\rm Born} + {\cal M}^\gamma_{\rm rad} +
{\cal M}^Z_{\rm Born} + {\cal M}^Z_{\rm rad}\right|^2
+  \left| {\cal M}^\gamma_{\rm Born} + {\cal M}^Z_{\rm Born} \right|^2
\kappa_{\rm soft}^{eq},
\end{equation}
where ${\cal M}^\gamma_{\rm rad}$ and ${\cal M}^Z_{\rm rad}$ denote the
one-loop radiative corrections. Note that at this stage, our
expression (\ref{e1.15b}) also includes the purely QED radiative corrections
${\cal M}^\gamma_{\rm rad}$ to the Born amplitude ${\cal M}^\gamma_{\rm
Born}$. ${\cal M}^\gamma_{\rm rad}$ includes vertex and $\gamma\gamma$
box diagrams of order ${\cal O}(\alpha^2)$.

For the PV contribution, it is convenient to work with the amplitude
${\cal M}_{\rm PV}^{eq}$ defined in Eq.~(\ref{e1.1}), from which we can
write the relevant PV interference term as
\begin{eqnarray}
d\sigma_{\rm PV,tot}^{eq}&\sim& 2\Re\biggl\{
\left({\cal M}^\gamma_{\rm Born}\right)^*
\left({\cal M}_{\rm PV,tree}^{eq} + {\cal M}_{\rm PV,rad}^{eq} + 
{\cal M}_{\rm PV,tree}^{eq} \half\kappa_{\rm soft}^{eq}\right) \nonumber\\
&&+ \left({\cal M}_{\rm PV,tree}^{eq}\right)^*
\left({\cal M}^\gamma_{\rm rad} + 
{\cal M}^\gamma_{\rm Born} \half\kappa_{\rm soft}^{eq}\right)\biggr\}.
\label{e1.13}
\end{eqnarray}

At this point we note that the second term of Eq.~(\ref{e1.13})  involves
a PV interference term between ${\cal M}^\gamma_{\rm rad}$ and ${\cal
M}_{\rm PV,tree}^{eq}$, plus the soft photon emission contribution needed
to cancel the IR divergences in ${\cal M}^\gamma_{\rm rad}$. In principal,
one should use the full expression (\ref{e1.13}) to evaluate both the QED
and weak radiative corrections on the same footing, since they contribute
to the PV asymmetry at the same order. In practice, this has not been
done. We take the position here that experimental analyses have already
accounted for the QED radiative corrections, and to include such effects
again here would be double counting. Accordingly, we drop the second term
in Eq.~(\ref{e1.13}) from further consideration at this time. 

Using the general structure of the PV amplitude in Eq.~(\ref{e1.1}), we
can rewrite the first term in Eq.~(\ref{e1.13}) as
\begin{eqnarray}
d\sigma_{\rm PV,tot}^{eq}&\sim& 2\Re\biggl\{\left(
{\cal M}_{\rm Born}^\gamma\right)^*\biggl[\left( 1+R_A^q+\half\kappa
_{\rm soft}^{eq}\right) A_{\rm tree}^{eq} J_{VA}^{eq} \nonumber\\
&&+\left(1+
R_V^q+\half\kappa_{\rm soft}^{eq}\right) B_{\rm tree}^{eq}
J_{AV}^{eq}\biggr]\biggr\},
\end{eqnarray}
where current products $J_{VA}^{eq}$ and $J_{AV}^{eq}$ are defined in
Eq.~(\ref{e1.1}).

Evaluating Eq.~(\ref{e1.15}), we have
\begin{eqnarray}
\kappa_{\rm soft}^{eq}&=&-{\alpha \over 2\pi^2}\biggl[
2m_e^2I\left( p_1,p_1\right)
-\left(2m_e^2-t\right) I\left( p_1,p_3\right)
+2Q_q^2m_q^2I\left( p_2,p_2\right) \nonumber\\
&& -Q_q^2\left( 2m_q^2-t\right) I\left( p_2,p_4\right) 
-2Q_q\left( u-m_e^2-m_q^2\right) I\left( p_1,p_4\right) \label{e1.16}\\
&& -2Q_q\left(s-m_e^2-m_q^2\right) I\left( p_1,p_2\right) 
\biggr], \nonumber 
\end{eqnarray}
with the soft-photon emission integral $I\left( p_i,p_j\right)$ defined as
\begin{equation}
I\left(p_i,p_j\right) =\int_{\left| \bbox{k}\right| \leq \Delta \varepsilon}
{d^3k\over 2\sqrt{\bbox{k}^2+\lambda ^2}}\,{1\over \left( p_i\cdot k\right)
\left(p_j\cdot k\right) }.
\label{e1.17}
\end{equation}
These integrals generally have been worked out in Ref.~\cite{HV79}.
For electron-quark scattering in the CM frame, we have (see also
Ref.~\cite{Hah97})
\begin{eqnarray}
I\left( p_i,p_j\right) &=&{2\pi \alpha_{ij} \over \alpha_{ij}^2m_i^2-m_j^2}
\biggl[ 
{1\over 2} \ln \left( {\alpha_{ij}^2 m_i^2\over m_j^2}\right) \ln \left( {
4\Delta \varepsilon^2\over \lambda ^2}\right) +{1\over 4} \ln^2\left(
{E_i-p_i\over E_i+p_i}
\right) -{1\over 4}\ln^2\left( {E_j-p_j\over E_j+p_j}\right) \nonumber\\ 
&&+\Li_2\left( 1-{\alpha_{ij}\over v_{ij}}\left( E_i+p_i\right) \right)
+\Li_2\left( 1-{\alpha_{ij}\over v_{ij}}\left( E_i-p_i\right) \right)
\label{e1.18} \\ 
&& -\Li_2\left( 1-{1\over v_{ij}}\left( E_j+p_j\right) \right) -\Li_2\left( 1-
{1\over v_{ij}}\left( E_j-p_j\right) \right) 
\biggr],  \nonumber
\end{eqnarray}
where $v_{ij}=(\alpha_{ij}^2m_i^2-m_j^2)/(2\left( \alpha
_{ij}E_i-E_j\right))$, and $\left( E_i,p_i\right) $ are the energy and
momentum in the CM system. The parameters for different values of $i,j$
are given in Table~\ref{table1}. $\Li_2$ is the dilogarithm function
\begin{equation}
\Li_2(z) = \int_z^0 dt\, {\ln (1-t)\over t}.
\end{equation}

\begin{table}[ht]
\caption{Soft photon emission integral parameters of Eq.~(\ref{e1.18}).}
\rule{0in}{2ex}
\begin{center}
\begin{tabular}{ccccc}
$i$ & $j$ & $m_i$ & $m_j$ & $\alpha_{ij}$\\ 
\tableline
1 & 1 & $m_e$ & $m_e$ & 1\\
2 & 2 & $m_q$ & $m_q$ & 1\\
1 & 3 & $m_e$ & $m_e$ & $1-{ t\over 2m_e^2}+{\sqrt{t^2-4t m_e^2}\over
2m_e^2}$\\
2 & 4 & $m_q$ & $m_q$ & $1-{ t\over 2m_q^2}+{\sqrt{t^2-4t m_q^2}\over
2m_q^2}$\\
1 & 4 & $m_e$ & $m_q$ & 
${m_e^2+m_q^2-u+\sqrt{\left( u-m_e^2-m_q^2\right)
^2-4m_e^2m_q^2}\over 2m_e^2}$\\
1 & 2 & $m_e$ & $m_q$ & 
${s-m_e^2-m_q^2+\sqrt{\left( m_e^2+m_q^2-s\right)
^2-4m_e^2m_q^2}\over 2m_e^2}$\\
\end{tabular}
\end{center}
\label{table1}
\end{table}

The IR divergent ($\ln \lambda$) terms in the elastic and inelastic
contributions to the radiative correction cancel exactly. In
particular, if we define the modified radiative correction factors
\begin{mathletters}
\begin{eqnarray}
{\tilde R}_V^q &\equiv& R_V^q + \half\kappa_{\rm soft}^{eq}, \\
{\tilde R}_A^q &\equiv& R_A^q + \half\kappa_{\rm soft}^{eq}.
\end{eqnarray}
\end{mathletters}
then ${\tilde R}_V^q$ and ${\tilde R}_A^q$ are the radiative corrections
to electron-quark scattering free of IR divergences. This was confirmed
numerically in our calculation by varying the parameter $\lambda$ over
several orders of magnitude, with no numerically significant change in
$\tilde R$. 

There remains a weak logarithmic dependence on $\Delta \varepsilon$ of the
form $\ln(\Delta\varepsilon/m)$. This dependence will be cancelled if hard
photon bremsstrahlung is taken into account. This requires knowledge of
particular experimental details such as detector geometry, energy
resolution, phase space cuts, etc. This is beyond the scope of this paper,
and we have chosen to simply set the scale for these effects by choosing
an energy resolution $\Delta E$ corresponding to the parameters of the
SAMPLE detector. We leave a more complete treatment of the hard
bremsstrahlung corrections, as well as the simultaneous treatment of both
QED and weak radiative corrections, as a future project.

\section{Numerical results}

The parameters of the Standard Model are taken from Ref.~\cite{PDG00}, and
are given in Table~\ref{table2}. As discussed by Hollik \cite{Hol90}, the
masses of the light quarks are regarded as parameters, and are adjusted to
reproduce the results of a dispersion analysis of the experimentally
measured hadronic vacuum polarization.

\begin{table}[ht]
\caption{Standard model parameters used in this calculation.}
\rule{0in}{2ex}
\begin{center}
\begin{tabular}{cccc}
Quantity& Value& Quantity& Value \\
\tableline
$m_u$ & 47.~MeV & $m_e$ & 0.51100~MeV\\
$m_d$ & 47.~MeV & $m_\mu$ & 105.66~MeV\\
$m_s$ & 150.~MeV & $m_\tau$ & 1777.0~MeV\\
$m_c$ & 1.25~GeV & $M_Z$ & 91.1882~GeV\\
$m_b$ & 4.2~GeV & $M_W$ & 80.419~GeV\\
$m_t$ & 174.3~GeV & $M_H$ & 100.~GeV\\
\end{tabular}
\end{center}
\label{table2}
\end{table}

We have chosen the SAMPLE kinematics, corresponding to $E_{e,{\rm
lab}}=194$~MeV and an average backward angle $\theta =146.2^\circ$.  
Radiative correction results for the $V(e)\times A(q)$ and $A(e)\times
V(q)$ current interactions are given separately in the Table~\ref{table3}
for $u$, $d$, and $s$ quarks.

The contributions from self-energy, triangle, and box classes of diagrams
are shown for loops which do not involve photons. The self-energy (SE)
diagrams give the dominant contribution to radiative corrections in most
of the cases.  Triangle and box diagrams that involve photon loops are
shown as a combined result in column 5 because they are separately
infrared divergent. In the combined result, the infrared divergence is
cancelled by the bremsstrahlung corrections for soft-photon emission. The
corrections show a weak logarithmic dependence on the resolution of the
detector, $\Delta E$, which we have taken to be 120~MeV in this table.
This corresponds roughly to the parameters of the SAMPLE detector, which
detects all electrons above the \v{C}erenkov threshold of 20~MeV
\cite{SAMPLE}. 

\begin{table}[ht]
\caption{Radiative corrections for one-quark currents at SAMPLE~I kinematics,
($E_{\rm lab}=194$~MeV, $\theta=146.2^\circ$, $\Delta E=120$~MeV).}
\rule{0in}{2ex}
\begin{center}
\begin{tabular}{lddddd}
$R_{VA}^{ff'}$ & Self Energy & Vertex (no $\gamma$)& Box (no $\gamma$) &
Vertex+Box (with $\gamma$)& Total \\ 
\tableline
$ue$ & -0.059 & -0.015 &  0.027 &  0.019 & -0.028 \\ 
$de$ & -0.025 &  0.026 &  0.002 &  0.004 &  0.008 \\ 
$se$ & -0.025 &  0.026 &  0.002 &  0.015 &  0.017 \\
$eu$ & -0.278 & -0.224 &  0.096 &  0.114 & -0.292 \\ 
$ed$ & -0.278 & -0.164 &  0.022 & -0.083 & -0.503 \\ 
$es$ & -0.293 & -0.198 &  0.022 & -0.081 & -0.550 \\ 
\end{tabular}
\end{center}
\label{table3}
\end{table}

The calculated one-quark radiative corrections were combined to form
hadronic vector and axial vector corrections. These are shown in
Table~\ref{table4} for the isoscalar and isovector representations
$R_V^{T=0}$, $R_V^{T=1}$, $R_A^{T=0}$, and $R_A^{T=1}$. Also shown are the
appropriate linear combinations of isoscalar and isovector corrections for
the proton ($R_V^p$ and $R_A^p$) and the neutron ($R_V^n$ and $R_A^n)$. In
addition, radiative corrections for the new SAMPLE kinematics, with
$E_{\rm lab}=120$~MeV and denoted SAMPLE~II, are also shown. Differences
with the SAMPLE~I results are mainly due to the bremsstrahlung
corrections.

\begin{table}[ht]
\caption{Radiative corrections for nucleon currents at SAMPLE~I kinematics,
($E_{\rm lab}=194$~MeV, $\theta=146.2^\circ$, $\Delta E=120$~MeV),
and SAMPLE~II kinematics
($E_{\rm lab}=120$~MeV, $\theta=146.2^\circ$, $\Delta E=77$~MeV).}
\rule{0in}{2ex}
\begin{center}
\begin{tabular}{llddddd}
& & T=0 & T=1 & isosinglet & p & n\\ 
\tableline
SAMPLE I  & $R_V$ &  0.057 & -0.005 &  0.029 & -0.252 &  0.022 \\ 
          & $R_A$ & -0.210 & -0.398 & -0.760 & -0.608 & -0.187 \\
\cline{2-7}
SAMPLE II & $R_V$ &  0.053 & -0.012 &  0.024 & -0.271 &  0.017 \\ 
          & $R_A$ & -0.216 & -0.408 & -0.781 & -0.624 & -0.192 \\
\end{tabular}
\end{center}
\label{table4}
\end{table}

We point out again that these values are calculated in the CDR scheme, and
also include a bremsstrahlung contribution. Values calculated in the
$\overline{\rm MS}$ scheme tend to be smaller \cite{Mus94}, however the
tree level amplitudes also differ between the two schemes due to the
different definitions of $\sin^2{\theta_W}$.

It is interesting to examine the dependencies of the total radiative
corrections $R_A^{T=1}$ and $R_V^p$ on various kinematic and Standard
Model parameters. These two corrections are large because the tree level
amplitudes are suppressed by $(1-4\sin^2\theta_W)$. We have found only a
weak dependence of the radiative corrections on scattering angle $\theta$
and electron energy $E_{\rm lab}$.

\begin{figure}[ht]
\begin{center}
\epsfig{figure=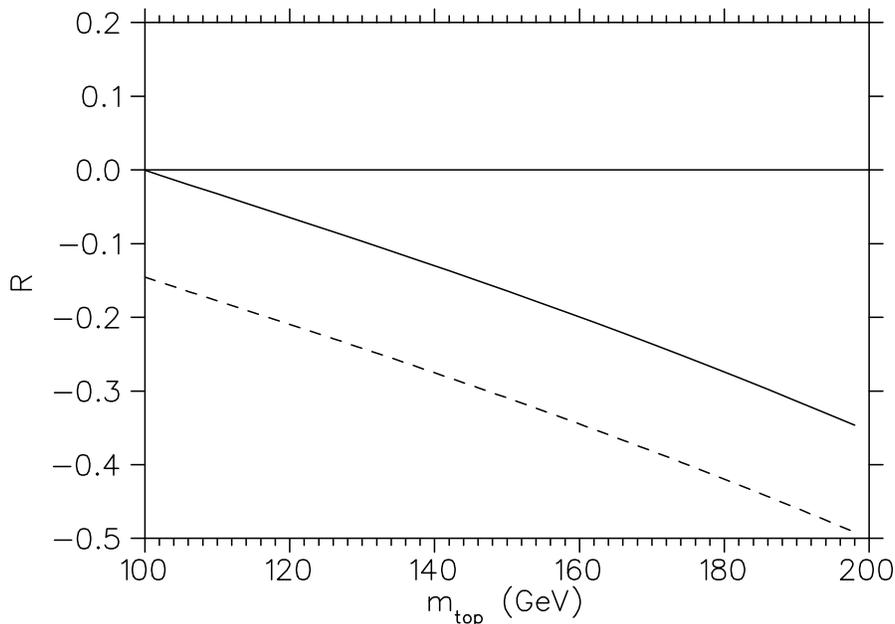,width=0.5\textwidth,
angle=90,clip=0}
\caption {$R_{A}^{T=1}$ (dashed line) and $R_{V}^{P}$ (solid line) as
a function of the mass of the top quark (in GeV).}
\end{center}
\end{figure}

Figure~4 shows the strong dependence of radiative corrections on the mass
of the top quark. It also facilitates comparison with previous work
\cite{MH90}, which used $m_t=120$~MeV. This strong dependence on $m_t$
arises from the large mass splitting within the top-bottom fermion
doublet. The corrections show only a slight dependence on the Higgs mass
$M_H$, which we have not shown. In Fig.~5 we show the (logarithmic)
dependence of $R_V^p$ and $R_A^{T=1}$ on the the photon detection
parameter $\Delta E$.

\begin{figure}[ht]
\begin{center}
\epsfig{figure=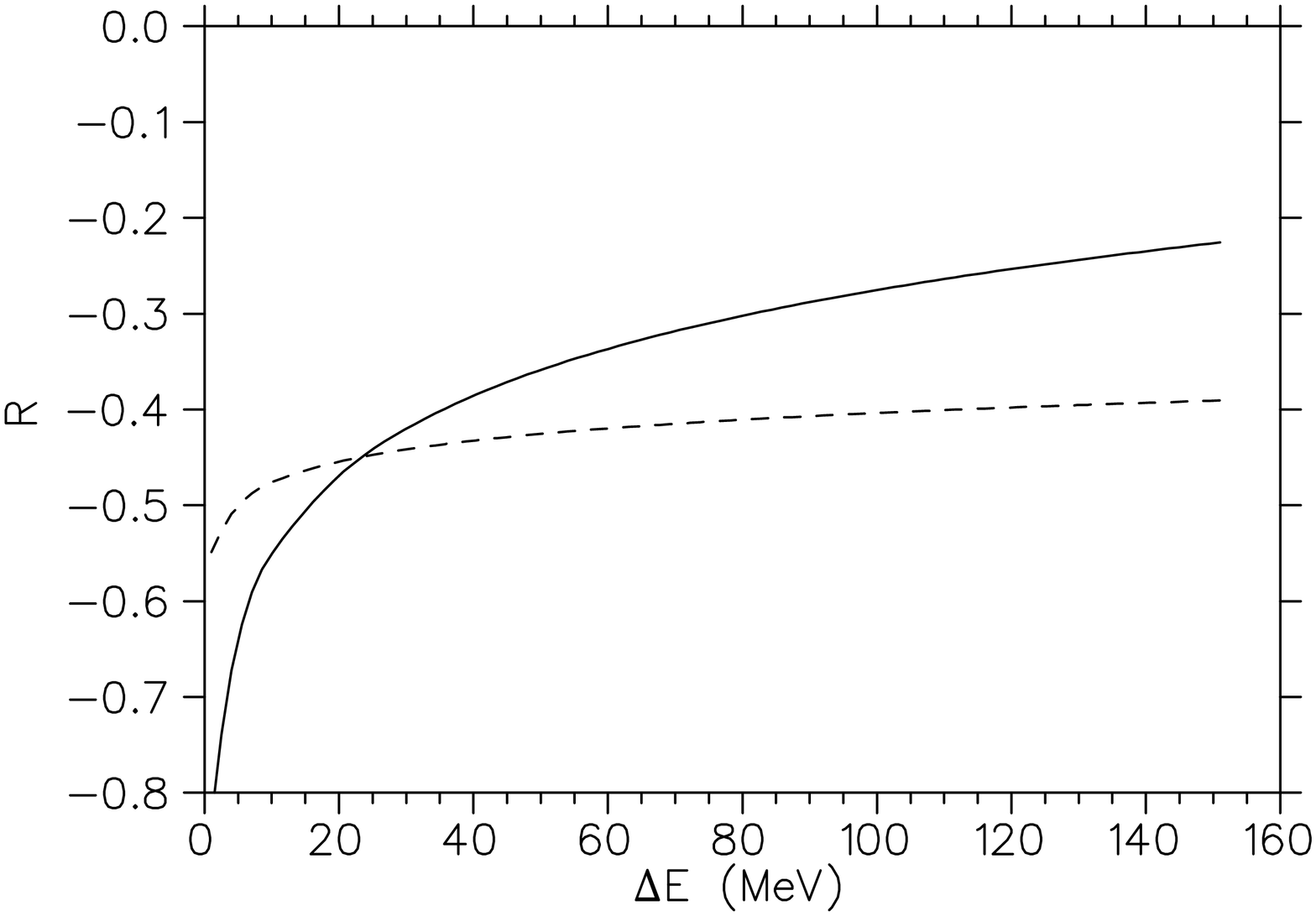,width=0.5\textwidth,
angle=90,clip=0}
\caption {$R_{A}^{T=1}$ (dashed line) and $R_{V}^{P}$ (solid line) as
a function of the photon detection parameter $\Delta E$ (in MeV).}
\end{center}
\end{figure}

\section{Experiment}

The first experiment to measure the weak neutral magnetic form factor
$G_M ^s$ of the nucleon is the SAMPLE experiment. The details of the
techniques employed in this experiment are available in
Refs.~\cite{BH01,Bei96,SAMPLE}. The experiment involves scattering 194
MeV polarized electrons from protons and deuterium. The scattered
electrons are detected at 4 backward angles $138^{\circ} < \theta <
160^{\circ}$, which results in an average $Q^2\simeq 0.1({\rm
GeV/c})^2$.

The measured proton and deuteron asymmetries for $Q^2=0.1$~(GeV/c)$^2$ and
$\theta_{\rm avg}=146.2^{\circ}$ are (in ppm):
\begin{mathletters}
\begin{eqnarray}
A_p^{\rm exp }&=&-4.92\pm 0.61\pm 0.73,\\
A_d^{\rm exp }&=&-6.97\pm 0.64\pm 0.55.
\end{eqnarray}
\end{mathletters}
The first uncertainty is statistical while the second is systematic. In
a subsequent analysis, Spayde \cite{Spa01} has reexamined the data, and
finds two significant corrections. The first arises from a different
treatment of the electromagnetic radiative corrections, while the
second is a correction for the fraction of the signal due to pions.
Each correction increases $A_p$ and $A_d$ by about 4 percent, resulting
in quoted asymmetries of
\begin{mathletters}
\begin{eqnarray}
A_p^{\rm exp }&=&-5.61\pm 0.67\pm 0.88,\\
A_d^{\rm exp }&=&-7.28\pm 0.68\pm 0.75.
\end{eqnarray}
\label{e3.2}
\end{mathletters}

Our theoretical results for the one quark contributions given in
Table~\ref{table4} need to be supplemented by the many quark
contributions. These have been evaluated using chiral perturbation theory
by Zhu {\it et al.} \cite{Zhu00}, who found small, predominantly
isovector, contributions.  Their isovector contribution can be
parameterized as $R_A^{T=1}=-0.06\pm 0.24$, while the isoscalar is
$R_A^{T=0}=-0.01\pm 0.14$. The quoted uncertainties are large, and
dominate the overall uncertainties in the theoretical analysis, as
described below.

Our error analysis for the theoretical asymmetry takes into account
uncertainties in the measured electromagnetic form factors, in the axial
contributions $G_A^{(8)}$ and $G_A^s$, and in the radiative corrections.
The largest source of uncertainty in the one quark radiative corrections
is in the soft photon approximation for the bremsstrahlung contributions.
We have allowed for a rather generous error estimate by evaluating these
corrections for $\Delta E = 60$~MeV, which is half the value used in
Table~\ref{table4}.

Our theoretical asymmetry for the SAMPLE kinematics, including all
radiative corrections, and adding errors in quadrature, can be written
to show the explicit dependence on the strange form factors:
\begin{mathletters}
\begin{eqnarray}
A_p^{\rm th}&=&(-7.29\pm 0.65)+(3.61\pm 0.07) G_M^s(0.1) + 1.94 G_E^s(0.1),\\
A_d^{\rm th}&=&(-8.74\pm 0.89)+(0.82\pm 0.05) G_M^s(0.1) + 1.46 G_E^s(0.1).
\end{eqnarray}
\end{mathletters}
For no strange quark contribution, the measured asymmetries are smaller
than the theoretical ones. The discrepancy between the theoretical and
experimental values for $A_p$ and $A_d$ are about the same, whereas the
coefficients of the dominant $G_M^s$ term are considerably different.

In our analysis we have used the recent nucleon electromagnetic form
factors determined by Brash {\it et al.}~\cite{Bra02}. At $Q^2=0.1$~GeV,
these nucleon form factors tend to be smaller than in the Galster
parameterization, thereby exacerbating the discrepancy between theoretical
and experimental values noted in previous work \cite{SAMPLE,Spa01}. The
axial form factors are taken to be a simple dipole form, following
Refs.~\cite{Mus94,SAMPLE}.

The contribution from $G_E^s$ for these kinematic is expected to be small.
For the range of values of the strange radius $r_s^2=\pm 0.22$ found in
the literature, we expect $G_E^s(0.1)$ to be in the range $\mp 0.10$. For
comparison, the neutron electric form factor has the value
$G_E^n(0.1)=0.036$.  HAPPEX \cite{HAPPEX} has measured the asymmetry $A_p$
at $\theta=12.3^\circ$ and $Q^2=0.477$. They find the linear combination
$(G_E^s+0.392 G_M^s)=0.025\pm 0.020 \pm 0.014$, which would seem to favor
a positive value of $G_E^s$. A measurement at $Q^2=0.1$ is underway, and
could shed some light on the SAMPLE result.

Because the largest overall uncertainty in the theoretical calculations is
from $R_A^{T=1}$, it is useful to follow the analysis of the SAMPLE papers
\cite{SAMPLE} and isolate this term explicitly. In this case, we have
\begin{mathletters}
\begin{eqnarray}
A_p^{\rm th}&=&(-5.93\pm 0.27)+(3.61\pm 0.07) G_M^s(0.1)
+ (2.34\pm 0.03) G_A^{Z,T=1}(0.1),\\
A_d^{\rm th}&=&(-7.10\pm 0.40)+(0.82\pm 0.05) G_M^s(0.1)
+ (2.83\pm 0.04) G_A^{Z,T=1}(0.1).
\end{eqnarray}
\label{e3.3}
\end{mathletters}
We have dropped the $G_E^s$ terms, and have added (in quadrature) an
additional uncertainty $\pm 0.18$ to $A_p$ and $\pm 0.14$ to $A_d$. The
isoscalar terms in $G_A^Z$, shown in Eq.~(\ref{e2.6}) have been absorbed
into the constant term. The axial isovector contribution is
\begin{equation}
G_A^{Z,T=1}(0.1) = - (1+R_A^{T=1}) G_A^{T=1}(0.1),
\end{equation}
which has the theoretically expected value $G_A^{Z,T=1}(0.1)=-0.58\pm
0.27$.  This includes the large effect of the one quark axial correction
($R_A^{T=1}=-0.40\pm 0.02$), and the highly uncertain many quark
contribution ($R_A^{T=1}=-0.06\pm 0.24$) of Ref.~\cite{Zhu00}.

Because the coefficients of $G_A^{Z,T=1}$ in Eqs.~(\ref{e3.3}) are roughly
equal, changes in $G_A^{Z,T=1}$ will shift $A_p$ and $A_d$ by about the
same amount. This is consistent with the uniform discrepancy between
theory and experiment in $A_p$ and $A_d$ referred to above, and so it is
useful to try to fit the data in this way. However, there may be other
explanations for such a uniform discrepancy, including other physics, or
systematic errors in the experimental analysis of the kind determined by
Spayde \cite{Spa01}. Hence we regard this as simply a device for
characterizing the data, and indicating possible directions for
explanation.

Solving the pair of equations (\ref{e3.3}) for $G_M^s$ and $G_A^{Z,T=1}$,
and using the measured asymmetries from Eq.~(\ref{e3.2}), we find
\begin{mathletters}
\begin{eqnarray}
G_M^s(Q^2=0.1) &=& 0.16 \pm 0.50 ,\\
G_A^{Z,T=1}(Q^2=0.1) &=& -0.11 \pm 0.49.
\end{eqnarray}
\end{mathletters}
Combining the errors in the extracted value of $G_A^Z$ and the
theoretical value quoted above, the difference is about 1 $\sigma$.

There is some uncertainty about how to extrapolate $G_M^s(Q^2=0.1)$ to
$\mu_s=G_M^s(Q^2=0)$. A simple dipole form consistent with the proton and
neutron magnetic electromagnetic form factors would give $\mu_s=0.21\pm
0.65$. Using the extrapolation model proposed by Hemmert {\it et al.}
\cite{HMS98}, one instead finds a value of the strange magnetic moment
$\mu_s = 0.04 \pm 0.50 \pm 0.07$, Here the last value of the error was
introduced in the extrapolation. A negative value of $\mu_s$ is consistent
with expectations from a number of theoretical models
\cite{Jaf89,DLW98,LT00,MD01}. In particular, recent lattice QCD
calculations find $\mu_s=-0.16\pm 0.18$ \cite{LT00} and $\mu_s=-0.28\pm
0.10$ \cite{MD01}. By contrast, a recent SU(3) chiral quark-soliton model
gives $\mu_s=0.074$ to $0.115$ \cite{SKG02}. 

We have repeated the above analysis for the SAMPLE~II kinematics
($E=120$~MeV). This experiment is currently being analyzed. We find
\begin{mathletters}
\begin{eqnarray}
A_p^{\rm th}&=&(-2.13\pm 0.08)+(1.07\pm 0.02) G_M^s(0.043)
+ (1.06\pm 0.01) G_A^{Z,T=1}(0.043),\\
A_d^{\rm th}&=&(-2.62\pm 0.13)+(0.26\pm 0.02) G_M^s(0.043)
+ (1.35\pm 0.01) G_A^{Z,T=1}(0.043),
\end{eqnarray}
\end{mathletters}
with the theoretically expected value $G_A^{Z,T=1}(0.043)=-0.62\pm0.28$.
The overall asymmetries are smaller, and the uncertainties enter somewhat
differently than in the SAMPLE~I kinematics. This should help in
determining the role of $G_A^{Z,T=1}$ in the asymmetries.

\section{Summary}

We have reexamined the one-loop radiative corrections to the one-quark
amplitudes in the CDR scheme. Our calculations include an evaluation of
bremsstrahlung corrections in the soft photon approximation. The
bremsstrahlung corrections cancel infrared divergences in the one-loop
diagrams. The soft photon approximation has a logarithmic dependence on
detection threshold. The sensitivity to this parameter will be reduced if
hard bremsstrahlung photons are included, but that requires knowledge of
the particular experimental setup.

Although we find that $G_A^{Z,T=1}$ is highly suppressed, there is no
indication from the present calculation that it is suppressed by the
magnitude indicated by the SAMPLE analysis. A contribution from the
isosinglet strange axial form factor $G_A^s$ is unlikely to resolve the
problem, as we find it is highly suppressed by the isosinglet radiative
corrections $R_A^{(0)}=-0.760$.  The many quark corrections calculated by
Zhu {\it et al.} \cite{Zhu00} were numerically small, but given the large
uncertainties, this is one avenue for further theoretical work.  Further
experimental measurements by the HAPPEX collaboration at
$Q^2=0.1$~GeV$^2$, as well as a new measurement by the SAMPLE group at the
smaller electron energy 120 MeV, should also shed light on this situation. 

\acknowledgements
This work was supported in part by the Natural Sciences and Engineering
Research Council of Canada, and by the US Department of Energy. The
Southeastern Universities Research Association (SURA) operates the Thomas
Jefferson National Accelerator Facility under DOE contract
DE-AC05-84ER40150. PGB would like to acknowledge and thank the theory
group at Jefferson Lab for support during a sabbatical leave, where part
of this work was completed.


\begin{references}

\bibitem{BM01} D.H.~Beck and R.D.~McKeown, Ann.~Rev.~Nucl.~Part.~Sci.
{\bf 51}, 189 (2001).
\bibitem{KM88} D.~Kaplan and A.~Manohar, Nucl.~Phys. {\bf B310}, 527 (1988).
\bibitem{Bec89} D.H.~Beck, Phys.~Rev.~D {\bf 39}, 3248 (1989).
\bibitem{Mus89} M.J.~Musolf, Ph.D.~thesis, Princeton University, (1989).
\bibitem{MH90} M.J.~Musolf and B.R.~Holstein, Phys.~Lett.~B {\bf 242},
461 (1990).
\bibitem{MH91} M.J.~Musolf and B.R.~Holstein, Phys.~Rev.~D {\bf 43},
2956 (1991).
\bibitem{Mus94} M.J.~Musolf, T.W.~Donnelly, J.~Dubach, S.J.~Pollock,
S.~Kowalski, and E.J.~Beise, Phys.~Rep. {\bf 239}, 1 (1994).
\bibitem{Zhu00} S.-L.~Zhu, S.J.~Puglia, B.R.~Holstein, and M.J.~Ramsey-Musolf,
Phys.~Rev.~D {\bf 62}, 033008 (2000).
\bibitem{HP99} T.~Hahn and M.~Perez-Victoria, Comp.~Phys.~Comm. {\bf 118},
153 (1999).
\bibitem{BH01} D.H.~Beck and B.R.~Holstein, hep-ph/0102053 (2001).
\bibitem{Bra02} E.J.~Brash, A.~Kozlov, Sh.~Li, and G.M.~Huber, Phys.~Rev.~C
{\bf 65} 051001, (2002).
\bibitem{Bei96} E.J.~Beise, {\it et al.}, Proc.\ SPIN96 Symp., 
Amsterdam, Sept. 1996, nucl-ex/9610011 (1996).
\bibitem{FJ01} B.W.~Filippone and X.~Ji, Adv.~Nucl.~Phys. {\bf 26}, 1 (2001).
\bibitem{Hah00} T.~Hahn, hep-ph/0005029 (2000).
\bibitem{Agu98} F.~del Aguila {\it et al.}, Nucl.~Phys. {\bf B537}, 561 (1999).
\bibitem{Sie79} W.~Siegel, Phys.~Lett.~B {\bf 84}, 193 (1979).
\bibitem{HS99} R.~Harlander and M.~Steinhauser, Prog.~Part.~Nucl.~Phys.
{\bf 43}, 167 (1999).
\bibitem{PV79} G.~Passarino and M.~Veltman, Nucl.~Phys. {\bf B160}, 151 (1979).
\bibitem{Hah97} T.~Hahn, Ph.D.~thesis, University of Karlsruhe, (1997).
\bibitem{BN37} F.~Bloch and A.~Nordsieck, Phys.~Rev. {\bf 37} (1937) 54.
\bibitem{MT00} L.C.~Maximon and J.A.~Tjon, Phys.~Rev.~C {\bf 62}, 054320
(2000).
\bibitem{HV79} G.~'t~Hooft and M.~Veltman, Nucl.~Phys. {\bf B153}, 365 (1979).
\bibitem{PDG00} Particle Data Group, D.E.~Groom {\it et al.}, Eur.~Phys.~J.
{\bf C15}, 1 (2000).
\bibitem{Hol90} W.F.~Hollik, Fortsch.~Phys. {\bf 38}, 165 (1990).
\bibitem{SAMPLE} B.A.~Mueller {\it et al.}, Phys.~Rev.~Lett. {\bf 78},3824
(1997);\\
D.T.~Spayde {\it et al.}, Phys.\ Rev.\ Lett. {\bf 84},1106 (2000).;\\
R.~Hasty {\it et al.}, Science {\bf 290}, 2117 (2000).
\bibitem{Spa01} D.T.~Spayde, Ph.D.~thesis, University of Maryland, (2001).
\bibitem{HAPPEX} K.A.~Aniol {\it et al.}, Phys.~Lett.~B {\bf 509}, 211 (2001).
\bibitem{HMS98} T.R.~Hemmert, U.-G.~Meissner, and S.~Steinberg, Phys.~Lett.~B
{\bf 437} 184, (1998).
\bibitem{Jaf89} R.L.~Jaffe, Phys.~Lett.~B {\bf 229}, 275 (1989).
\bibitem{DLW98} S.J.~Dong, K.F.~Liu, and A.G.~Williams, Phys.~Rev.~D
{\bf 58}, 074504 (1998).
\bibitem{LT00} D.B.~Leinweber and A.W.~Thomas, Phys.~Rev.~D {\bf 62},
074505 (2000).
\bibitem{MD01} N.~Mathur and S.-J.~Dong, Nucl.~Phys.~Proc.~Suppl. {\bf 94},
311 (2001).
\bibitem{SKG02} A.~Silva, H.-C.~Kim, and K.~Goeke, Phys.~Rev.~D {\bf 65},
014016 (2002).

\end{references}
\end{document}